\documentclass[amsfonts,a4paper,11pt]{article}

\usepackage{amsmath}
\usepackage{graphicx}
\usepackage{amsfonts}
\usepackage{amssymb}
\usepackage{bm}
\usepackage{xcolor}
\usepackage{color}

\usepackage[colorlinks=true, urlcolor=violet, linkcolor=blue,
  citecolor=red, hyperindex=true, linktocpage=true]{hyperref}

\newcommand{\bc}{\begin{center}}
\newcommand{\ec}{\end{center}}
\def\ba#1{\begin{array}{#1}\displaystyle}
\newcommand{\ea}{\end{array}}

\newcommand{\beq}{\begin{equation}}
\newcommand{\eeq}{\end{equation}}
\newcommand{\beqa}{\begin{eqnarray}}
\newcommand{\eeqa}{\end{eqnarray}}
\newcommand{\no}{\nonumber}
\newcommand{\n}{\nonumber\\}
\newcommand{\bi}{\begin{itemize}}
\newcommand{\ei}{\end{itemize}}

\def\lt#1{\left#1}
\def\rt#1{\right#1}
\def\t#1{\tilde{#1}}

\def\b#1{\bar{#1}}
\def\frc#1#2{\frac{#1}{#2}}

\newcommand{\vac}{{\rm vac}}
\newcommand{\bra}{\langle}
\newcommand{\ket}{\rangle}
\newcommand{\Z}{{\mathbb{Z}}}

\newcommand{\C}{{\mathbb{C}}}

\newcommand{\Or}{{\cal O}}

\newcommand{\varep}{\varepsilon}

\newcommand{\Tr}{{\rm Tr}}

\newcommand{\TT}{{\mathcal{T}}}

\def\eqref#1{(\ref{#1})}

\begin{document}

\begin{titlepage}

\begin{center}
{\Large {\bf Universal scaling of the logarithmic negativity in massive quantum field theory}}

\vspace{0.8cm} {\large \text{Olivier Blondeau-Fournier${}^{\circ}$, Olalla A. Castro-Alvaredo${}^{\heartsuit}$ and Benjamin Doyon${}^{\bullet}$}}

\vspace{0.2cm}
{{\small ${}^{\circ}$ ${}^{\bullet}$} Department of Mathematics, King's College London, Strand WC2R 2LS, UK}\\
{{\small ${}^{\heartsuit}$} Department of Mathematics, City University London, Northampton Square EC1V 0HB, UK }\\
\end{center}

\vspace{1cm}

\noindent We consider the logarithmic negativity, a measure of bipartite entanglement, in a general {unitary} 1+1-dimensional massive quantum field theory, not necessarily integrable. We compute the negativity between a finite region of length $r$ and an adjacent semi-infinite region, and that between two semi-infinite regions separated by a distance $r$. We show that the former saturates to a finite value, and that the latter tends to zero, as $r\to\infty$. We show that in both cases, the leading corrections are exponential decays in $r$ (described by modified Bessel functions) that are solely controlled by the mass spectrum of the model, independently of its scattering matrix. This implies that, like the entanglement entropy, the logarithmic negativity displays a very high level of universality, allowing one to extract {information about} the mass spectrum. Further, a study of sub-leading terms shows that, unlike the entanglement entropy, a large-$r$ analysis of the negativity allows for the detection of bound states.
\vfill

\noindent ${}^{\circ}$ olivier.blondeau-fournier@kcl.ac.uk\\
${}^{\heartsuit}$ o.castro-alvaredo@city.ac.uk\\
${}^{\bullet}$ benjamin.doyon@kcl.ac.uk \hfill \today

\end{titlepage}

\section{Introduction}
Entanglement is a fundamental property of quantum systems which relates to the outcomes of local measurements. In the context of 1+1 dimensional many body quantum systems, especially over the last 10 years, there has been growing interest in developing efficient
(theoretical) measures of entanglement as a means to extract valuable information about emergent properties of quantum states. Various measures of entanglement exist, see e.g.
\cite{bennet,Osterloh,Osborne,Barnum,Verstraete,VW,ple}, which occur in the context of quantum computing. Among these, a well-known and much studied measure of entanglement within many-body physics is the bipartite entanglement entropy \cite{bennet} (see \cite{specialissue} for various reviews). Another measure that has recently received some attention within many-body physics \cite{negativity1,negativity2} is the logarithmic negativity \cite{VW}, {which was shown to be an entanglement monotone in \cite{ple}}.

The entanglement entropy (EE) measures the amount of quantum
entanglement, in a pure quantum state, between the degrees of freedom associated to two sets of independent observables whose union is a complete set for the Hilbert space. Computations of the EE have revealed themselves as powerful tools especially in the study of quantum critical points, since the EE exhibits striking universal behaviours at and near such points. In 1+1 dimensions, the behaviour at criticality has been established analytically by employing conformal field theory (CFT) techniques, which have proven extremely effective in predicting the EE scaling laws in spatial bipartitions. The first of such results \cite{HolzheyLW94,Latorre1} established a fundamental relation between the logarithimic growth of the EE and the CFT central charge for connected regions in ground states of unitary models, later re-derived using different techniques \cite{Calabrese:2004eu,Calabrese:2005in}. Generalizations have included excited states \cite{german1, german2}, disconnected regions \cite{dis1,dis2,dis3,dis4,dis5} and non-unitary 
models \cite{BCDLR}. The first result beyond criticality \cite{Calabrese:2004eu} showed that the EE of infinite regions saturates to a value which logarithmically grows with the correlation length. In the full universal region near critical points, the EE is described by a universal scaling function related to correlation functions of branch point twist fields in massive quantum field theory (QFT) \cite{entropy}. Using this, a precise description of corrections to the saturation value has been found \cite{entropy,next}, showing that the leading system-size dependent correction displays highly universal and unexpected features controlled by the mass spectrum of the QFT.

\begin{figure}[h!]
\begin{center}
\includegraphics[width=7cm]{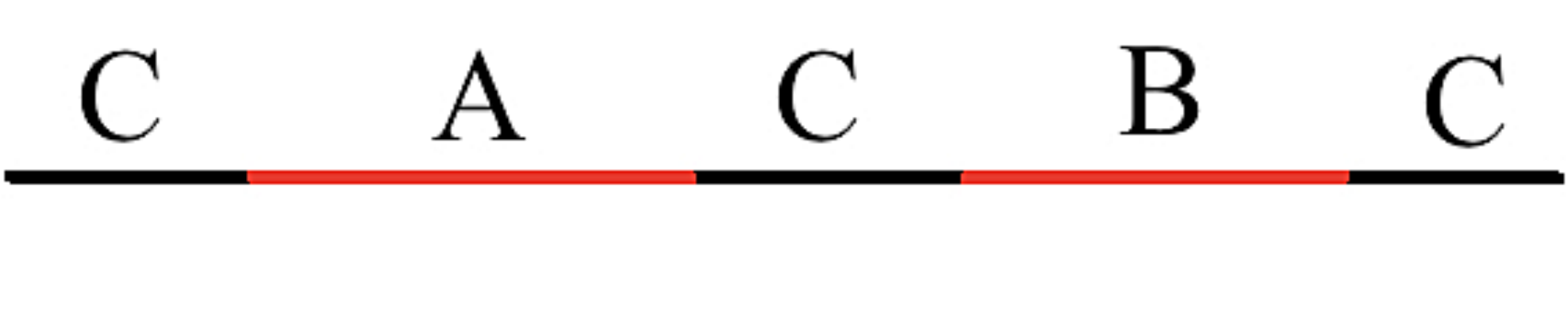} 
\includegraphics[width=7cm]{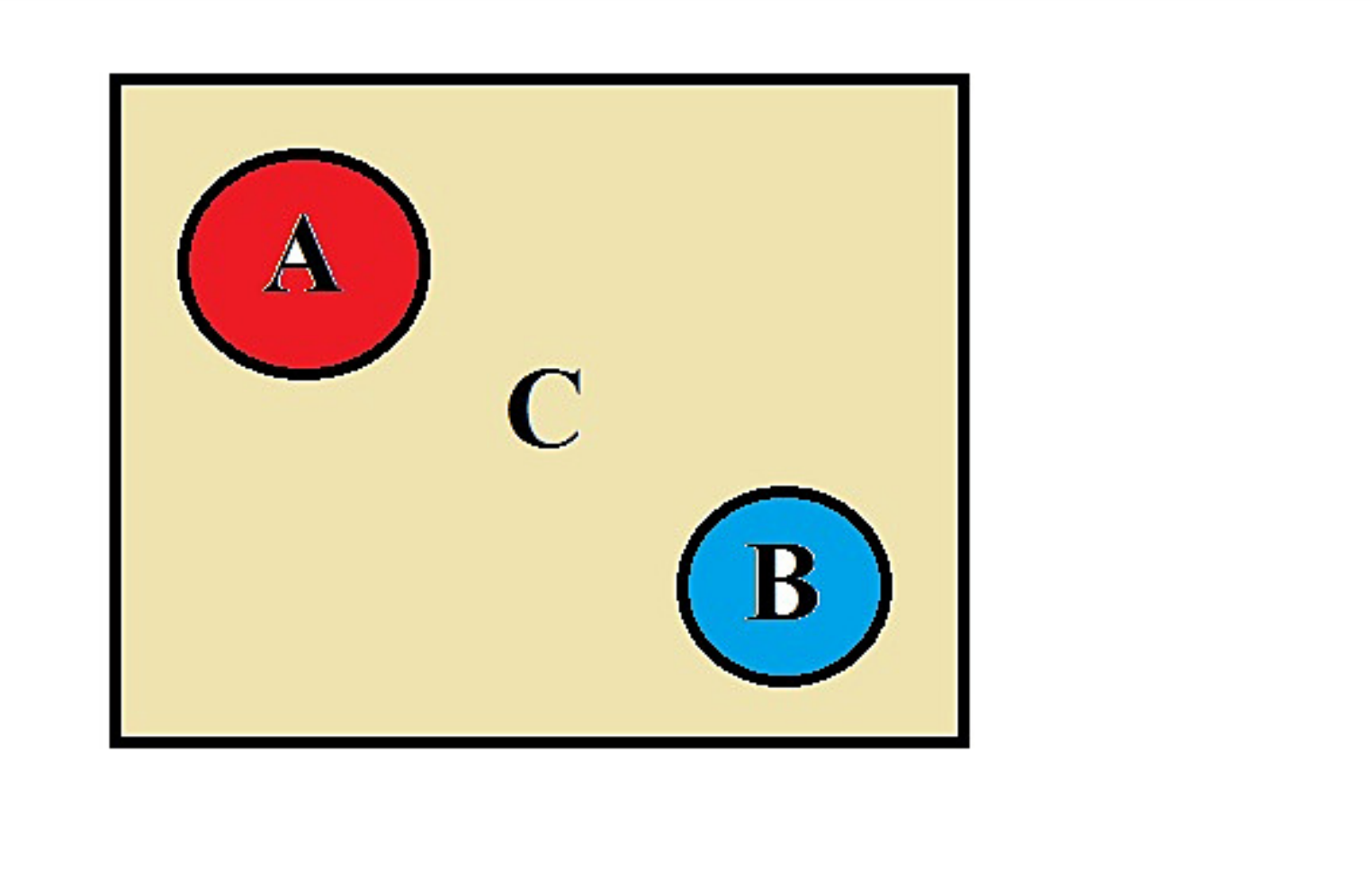} 
\end{center} 
\caption{Typical negativity configurations in 1+1- and 1+2-dimensions.}
\label{figneg}
\end{figure} 

The EE, however, is not a good measure of entanglement in {\em mixed states}. This, for instance, precludes its use for measuring the entanglement between {\rm non-complementary regions}, such as the regions $A$ and $B$ depicted in Fig.~1, since the reduced density matrix on the subspace corresponding to $A\cup B$ is generically mixed. An alternative measure of entanglement which is suited to treat mixed states is the logarithmic negativity introduced in \cite{VW,ple}. Powerful methods were developed in \cite{negativity1,negativity2} which allow its study in CFT using branch-point twist fields, much like in the case of the EE, and some universal results were found for critical points, again relating it to the CFT central charge. However, still relatively little is known about other universal features of the negativity. Detailed studies of free massless theories such as the compactified free Boson \cite{negativity2,numneg}, critical Ising spin chain \cite{negativity3} and free massless Fermion \cite{negativity4} have been carried out, as well as investigations of the negativity after a quantum quench \cite{negativity5,EZ1,negMB,WCR} and at finite temperature \cite{negativity6}. The negativity has also been recently considered within the holographic approach \cite{H1,H2}.

In this paper, we study the logarithmic negativity in general {unitary} 1+1 dimensional massive QFT. We consider two cases of the configuration presented in Fig.~1: (i) the regions $A$ and $B$ are adjacent, $B$ is semi-infinite and $A$ covers a finite but large distance; and (ii) both $A$ and $B$ are semi-infinite and separated by a finite but large distance. We obtain universal results for large-distance asymptotics purely in terms of QFT data. In particular, we find, like in the case of the EE, that the leading correction to the large distance constant (there is large-distance saturation in case (i), and vanishing in case (ii)) is completely determined by the mass spectrum of the model. Interestingly, unlike for the EE, we find that certain bound state data are involved in the first few sub-leading terms of the large-distance expansion in case (i).

This paper is organized as follows: in section 2 we provide a formal definition of the negativity and review its representation in terms of a four-point function of branch point twist fields \cite{negativity1,negativity2}. In section 3 we describe in- and out- states and the analytic properties of the matrix elements of branch-point twist fields in general massive 1+1 dimensional (replica) QFT \cite{next}. In section 4 we present a completely general computation of the leading correction to negativity saturation in massive 1+1 dimensional QFT for the cases of two adjacent and of two semi-infinite non-adjacent regions. In section 5 we present our conclusions and outlook. Specific examples are briefly discussed in appendix A.

\section{Logarithmic negativity in quantum field theory}

The logarithmic negativity provides a measure of the bipartite entanglement between two sets of observables in mixed states. Consider the geometric configuration presented in Fig.~1, with the whole system being in a pure state $|\psi\ket$. Tracing over the space corresponding to $C$, the logarithmic negativity of the resulting reduced density matrix provides a measure of the bipartite entanglement between regions $A$ and $B$. The precise definition goes as follows: let $\rho$ be the reduced density matrix $\rho=\Tr_{C}(|\psi\ket \bra \psi|)$ of the subsystem $A\cup B$ and consider the ``partial transpose'' of this density matrix, denoted by  $\rho^{T_B}$, with respect to the tensor factor corresponding to $B$. That is, let $|e_i^A\rangle$ and $|e_j^B\rangle$, for $i,j\in \mathbb{N}$, be basis state vectors on the Hilbert spaces associated to the degrees of freedom on $A$ and $B$ respectively, and denote $|e_i^Ae_j^B\rangle = |e_i^A\rangle_A \otimes |e_j^B\rangle_B$.  The matrix elements of $\rho^{T_B}$ are then given by the relations
$\bra e_i^{A}e_j^B|\rho^{T_B}|e_k^A e_l^B\ket=\bra e_i^{A}e_l^B|\rho|e_k^A e_j^B\ket$. Although this notion is basis-dependent (because the notion of transpose is not fundamental to a Hilbert space), it was shown in \cite{ple} that the logarithmic negativity, defined as
\beq \label{NegE}
\mathcal{E}=\log\Tr|\rho^{T_B}|
\eeq 
where $\Tr|\rho^{T_B}|$ represents the sum of the absolute values of the eigenvalues of $\rho^{T_B}$, is a good measure of entanglement in mixed states.

As is the case for the EE, QFT techniques are particularly well suited to the study of the negativity by using a replica approach. Such approach was described in detail in \cite{negativity1,negativity2} for CFT and is easily generalizable to {unitary} massive QFT and quantum chains. The description provided for CFT applies directly to the case we want to study.   
The idea is to compute 
\beq\label{NegEn}
\mathcal{E}[n]:=\log \Tr (\rho^{T_B})^n
\eeq
for positive integer values of $n$, and to ``analytically continue'' the resulting function of $n$. A key conclusion from the analysis presented in \cite{negativity1} is that the expression \eqref{NegEn} has different ``natural" analytic continuations $\mathcal{E}_e[n]$ and $\mathcal{E}_o[n]$ for different parities of $n$ (even and odd, respectively), and that the negativity \eqref{NegE} is obtained by taking the limit $n\to1$ from the analytic function $\mathcal{E}_e[n]$ in the {\em even} sector. That is:
\beq
\mathcal{E}=\lim_{n\rightarrow 1} \mathcal{E}_e[n].
\eeq
It was then found that $\Tr(\rho^{T_B})^{n}$ can be expressed, for integer values of $n$, in terms of correlation functions of local fields in a replica theory, composed of $n$ independant copies of the original model. The fields involved are the branch-point twist fields \cite{Kniz,orbifold,Calabrese:2004eu,entropy}, and the techniques and expressions obtained are similar to those used in the computation of the entanglement entropy \cite{Calabrese:2004eu,Calabrese:2005in,entropy}.

Branch-point twist fields are twist fields associated with elements of the permutation symmetry group of any replica ($n$-copy) model. Labelling the copies by $j=1,\ldots, n$, the branch-point twist field $\TT$ is associated with the cyclic permutation\footnote{Our convention is that the twist field $\TT$ has a branch cut to its right, and the continuity condition along this branch cut connects sheet number $j+1$ below the cut to sheet number $j$ above it.} $j\mapsto j+1\ {\rm mod}\,n$, and the ``anti-twist field'' $\t\TT$ is associated with the opposice permutation $j\mapsto j-1\ {\rm mod}\,n$ (these generate permutation cycles of length $n$). The locality properties of branch-point twist fields were described in detail in \cite{entropy}, leading to a set of consistency equations for their matrix elements in integrable 1+1-dimensional QFT, a part of which still holds in arbitrary massive QFT \cite{next}. In CFT, the conformal dimensions of both twist fields are known to be \cite{Kniz,orbifold, Calabrese:2004eu}
\beq 
\Delta_n=\frac{c}{24}\left(n-\frac{1}{n}\right),
\eeq 
where $c$ is the central charge of the CFT. In general, this formula holds for twist fields, in replica models, associated to any permutation element, where $n$ is the length of the permutation cycle generated. {We normalize branch-point twist fields in the usual conformal normalization, represented by the OPE
\beq 
	{\cal T}(0)\t{\cal T}(r)\sim r^{-4\Delta_n} {\bf 1} +
    \ldots.
\eeq 
Under this normalization, the CFT structure constant $C_{{\cal T}{\cal T}}^{{\cal T}^2}$ is unambiguous,
\beq 
	{\cal T}(0){\cal T}(r)\sim C_{{\cal T}{\cal T}}^{{\cal T}^2} \,{\cal T}^2(0) + \ldots
\eeq 
where ${\cal T}^2$ is the composite twist field associated with the permutation element $j\mapsto j+2\ {\rm mod}\,n$. The structure constant $C_{{\cal T}{\cal T}}^{{\cal T}^2}$ will play a role below.}

In terms of these fields, the negativity associated to a configuration such as that represented in Fig.~\ref{figneg} may be obtained in terms of a four point function of twist fields located at the boundary points of regions $A$ and $B$ \cite{negativity1}. Denoting the regions $A=[r_1,r_2]$ and $B=[r_3,r_4]$ with $r_1< r_2< r_3< r_4$, the quantity $\mathcal{E}[n]$ can be written in terms of the following four-points correlation function
\beq\label{ZnEn}
\mathcal{E}[n]= 8\Delta_n \log \varep + \log\lt({\,}_{n}\bra \text{vac}| \TT(r_1)\tilde{\TT}(r_2)\tilde{\TT}(r_3) \TT(r_4) |\text{vac}\ket_{n}\rt)
\eeq
where $|\vac\rangle_n$ is the $n$-tensor product of the vaccum state of the original theory, $|\vac\rangle_n=|\vac\rangle \otimes \cdots \otimes |\vac\rangle$. Here $\varep$ is a non-universal short-distance cutoff, and the second term is a universal scaling function.

The correlation function above can be very involved, even for CFTs (see \cite{negativity2} for a detailed study of the compactified massless free Boson), so that obtaining the analytic continuation $\mathcal{E}_e[n]$ and taking the limit $n\to1$ remains challenging but for the simplest of cases. In massive QFTs, even integrable ones, the evaluation of the four point function and its analytic continuation will be challenging. However, there are two special limiting cases which can 
be studied in much generality. We will consider those in this paper.

\subsection{First case: adjacent regions}

The first case, $\mathcal{E}^{\perp}$, corresponds to the negativity of {\it adjacent regions}, one of the regions being {\em semi-infinite}: the configuration for which $r_3\to r_2:=r$ and  $r_4\to\infty$ (we will choose $r_1=0$). We will show that, in this case, the negativity behaves as follows:
\beqa
	\mathcal{E}^{\perp} &\stackrel{m_1r\to0}\sim& \frc c4 \log (r/\varep) \qquad
	\n
	&\stackrel{m_1r\gg 1}=& -\frc c4 \log(m_1\varep) + \mathcal{E}_{\rm sat} -\frc{2}{3\sqrt{3}\pi}
    \sum_\alpha K_0(\sqrt{3}m_\alpha r)
    + O(e^{-Zm_1r})
    \label{main1}
\eeqa
where $Z>\sqrt{3}$ is some number that depends on the presence of particle--anti-particle bound states in the QFT model (if there are no such bound states, then $Z=2$). Here $\varep$ is a non-universal short-distance cutoff which may be conveniently chosen so as to ensure there are no $O(1)$ corrections to the first line of (\ref{main1}), $\{m_\alpha\}$ is the mass spectrum of the QFT with $m_1$ the lightest mass, and $K_0(x)$ is the modified Bessel function. The universal saturation constant is then given by
\beq\label{U}
	\mathcal{E}_{\rm sat} =
    2\log\lt(
    m_1^{\frc{c}{8}} {}_{\frc12}\bra\vac|\TT|\vac\ket_{\frc12}\rt) - \log(C_1),
\eeq
where $C_1$ is the CFT structure constant $C_{{\cal T}{\cal T}}^{{\cal T}^2}$ analytically continued from even number of copies $n$ to $n=1$, and where the expectation value is likewise understood as an analytic continuation in the number of copies. We remark that the saturation constant involves both a massive QFT quantity, the vacuum expectation value, and a pure CFT quantity, the structure constant $C_1$.

The limit $r_3\to r_2$ was studied in CFT in \cite{negativity1, negativity2} and is fully understood in this context. This limit is singular on the universal scaling function (as the fields $\tilde{\TT}(r_2)$ and $\tilde{\TT}(r_3)$ collide), a signal that the leading non-universal divergency is modified. The limit is of course well defined on the negativity itself (which has a non-universal part controlled by the miscrocopic theory), and after taking this limit, there is a new universal scaling function. This new scaling function can be obtained by using the composite field $\t{\TT}^2(r)$, associated to the permutation element $j\mapsto j+2\ {\rm mod}\,n$, in place of the product of fields $\tilde{\TT}(r_2)\tilde{\TT}(r_3)$. It is simple to see that for even values of $n$, this permutation element factorizes into permutation elements on even- and odd-numbered copies, each generating cycles of length $n/2$.
As a consequence, the composite twist fields factorizes, for even $n$, as a product of twist fields acting on even-numbered and odd-numbered sheets:
\beq\label{T2}
	\mathcal{T}^2 =
	\mathcal{T}|_{\mbox{\small $j$ even}}\otimes \mathcal{T}|_{\mbox{\small $j$ odd}} \qquad 
	\mbox{($n$ even)}
\eeq
In particular, for $n$ even its conformal dimension is
\beq\label{dimTT2}
	\Delta_{\TT^2}(n) = 2\Delta_{\frac{n}{2}} \qquad
    \mbox{($n$ even).}
\eeq
Note that by opposition, $\TT^2$ corresponds to a (non-sequential) cyclic permutation generating a full cycle of length $n$ when $n$ is odd. The field $\tilde{\TT}^2$ is, as usual, associated to the opposite permutation. In \eqref{ZnEn}, the dimension of the field $\TT^2$ for $n$ even implies that, in the configuration where $r_2=r_3$, the non-universal divergence of ${\mathcal{E}_e[n]}$ is instead of the form $4\lt(\Delta_n + \Delta_{\frac{n}{2}}\rt) \log \varep$, where $\varep$ is a (possibly different) short-distance cutoff.

We will further take the limit $r_4\to\infty$, so that one of the regions is of infinite length. In massive QFT, correlation functions factorize:
\[
	\lim_{m_1r_4\to\infty} {\,}_{n}\bra \text{vac}| \TT(r_1)\tilde{\TT}^2(r) \TT(r_4) |\text{vac}\ket_{n} =
    {}_n\bra\vac|\TT(r_1)\t\TT^2(r)|\vac\ket_n \,
    {}_n\bra\vac|\TT|\vac\ket_n.
\]
This means that the negativity saturates, as $r_4\to\infty$, to a value that depends on $r-r_1$: the quantity of entanglement between a semi-infinite region and a finite region is finite thanks to the finite correlation length. Here, the additional term $\log\lt({}_n\bra\vac|\TT(r_4)|\vac\ket_n\rt)$ can be absorbed into a re-definition of the short-distance cutoff $\varep$. The negativity is then related to the function (taking $r_1=0$)
\beq\label{Zn3p}
\mathcal{E}_e^{\perp}[n]= 4\lt(\Delta_n + \Delta_{\frac{n}{2}}\rt) \log \varep +\log\lt({\,}_{n}\bra \text{vac}| \TT(0)\tilde{\TT}^2(r)|\text{vac} \ket_{n}\rt)
\eeq
where the expectation value is analytically continued from positive even values of $n$. The universal information is encoded into the two-point function, and an analysis of the above, using $\Delta_{\frac{1}{2}} = -\frac{c}{16}$, will show that this leads to \eqref{main1} (upon a re-definition of $\varep$). 

\subsection{Second case: disjoint semi-infinite intervals}

Another interesting limiting case corresponds to the negativity $\mathcal{E}^{\dashv \, \vdash}$ of two {\it semi-infinite non-adjacent regions}, namely $r_{1}\to-\infty$, $r_4\to\infty$, and we will choose $r_2=0$, $r_3=r$. In this case, by factorization of correlation functions, the function $\mathcal{E}^{\dashv \, \vdash}_e[n]$ takes the form
\beqa  
\mathcal{E}^{\dashv \, \vdash}_e[n] = 8\Delta_n \log\varep +\log\left(
{\,}_{n}\bra \text{vac}| \tilde{\TT}(0)\tilde{\TT}(r)|\text{vac} \ket_{n}\right),\label{e2}
\eeqa  
where $\varep$ is some non-universal short-distance cutoff and again the correlation function is analytically continued from positive even values of $n$. We will show that this leads to the following:
\beqa
	\mathcal{E}^{\dashv \, \vdash} &\stackrel{m_1r\to0}\sim& -\frc c4 \log (m_1r) + \mathcal{E}_{\rm shift}
	\n
	&\stackrel{m_1r\gg 1}=&
	\frac{1}{2\pi^2} \sum_{\alpha} m_\alpha r\,\left[m_\alpha r K_0(m_\alpha r)^2+K_0(m_\alpha r)K_1(m_\alpha r)-m_\alpha r K_1(m_\alpha r)^2\right]
    + O(e^{-4m_1r})\n &&
    \label{sca2}
\eeqa
where the shift constant is
{\beq\label{Ud}
	\mathcal{E}_{\rm shift} =
    2\log\lt(
    m_1^{\frc{c}{8}} {}_{\frc12}\bra\vac|\TT|\vac\ket_{\frc12}\rt) + \log(C_1).
\eeq}
As before, {$C_1$ is the CFT structure constant $C_{{\cal T}{\cal T}}^{{\cal T}^2}$ analytically continued to $n=1$ from even copy numbers}, $\{m_\alpha\}$ is the mass spectrum of the QFT with $m_1$ the smallest mass, and $K_0(x),\,K_1(x)$ are modified Bessel functions. Note that both short- and large-distance behaviours are universal, and that the shift constant $\mathcal{E}_{\rm shift}$ is closely related to the saturation constant $\mathcal{E}_{\rm sat}$ given in \eqref{U}. {In particular, it is possible to extract from these universal quantities both the massive QFT and the pure CFT parts:
\beq\label{vacc}
	2\log\lt(
    m_1^{\frc{c}{8}} {}_{\frc12}\bra\vac|\TT|\vac\ket_{\frc12}\rt) =
    \frc{\mathcal{E}_{\rm shift}+\mathcal{E}_{\rm sat}}2,\quad
    \log(C_1) =
    \frc{\mathcal{E}_{\rm shift}-\mathcal{E}_{\rm sat}}2.
\eeq
}

\section{Scattering theory and form factors} \label{sectscatt}

Let us recall elements of the general scattering theory of massive QFT which will allow us to extract universal results for the negativity.

The space of states of massive QFT has two natural bases: the in-states and the out-states. These are scattering states, representing asymptotically free particles either in the far past or the far future. Both are labelled by the rapidity of the particles, and we denote by $|\theta_1\cdots \theta_\ell\ket_{\alpha_1,\ldots,\alpha_\ell}$ the state with asymptotic particles of rapidities $\theta_1,\ldots,\theta_\ell$ and internal quantum numbers $\alpha_1,\ldots,\alpha_\ell$, with the convention that it is an ``in" state if $\theta_1>\cdots>\theta_\ell$, and an ``out'' states in the opposite order. The conjugate is denoted as $|\theta_1\cdots \theta_\ell\ket^\dag_{\alpha_1,\ldots,\alpha_\ell} = {}_{\alpha_1,\ldots,\alpha_\ell}\bra\theta_1\cdots\theta_\ell|$.

The analytic continuation, in the space of rapidities, of (normalized) matrix elements of local fields of the form $\bra\vac|\Or|\theta_1\cdots \theta_\ell\ket_{\alpha_1,\ldots,\alpha_\ell}\, /\, \bra\vac|\Or|\vac\ket$ are referred to as form factors and denoted $F^{\Or}_{\alpha_1,\ldots,\alpha_\ell}(\theta_1\cdots \theta_\ell)$. These are in general complicated functions of the rapidities, but they possess certain analytic properties \cite{ELOP,Weisz,KW,smirnovbook}. The two-particle form factors $F^{\Or}_{\alpha_1,\alpha_2}(\theta_1,\theta_2)$ are particularly simple: by relativistic invariance, for spinless fields they only depend on the rapidity difference $\theta=\theta_1-\theta_2$, and as functions of this variable, they are analytic in the physical strip ${\rm Im}(\theta)\in[0,2\pi]$ except for possible bound state poles on ${\rm Re}(\theta)=0$, ${\rm Im}(\theta)\in(0,2\pi)$; the boundary values on $(0,\infty)$ and on $(2\pi i -\infty,2\pi i)$ are in-state matrix elements, and those on $(-\infty,0)$ and $(2\pi i ,2\pi i +\infty)$ are out-state matrix elements \cite{ELOP}.

Branch-point twist fields were studied within the general scattering theory in \cite{next}, following their study \cite{entropy} in integrable QFT. Recall that this twist field is defined on an $n$-copy replica model, whose Hilbert space is the $n$-times tensor power ${\cal H}^{\otimes n}$ of the original Hilbert space ${\cal H}$ of the QFT.  In such a replica model, any internal quantum number $\mu=(\alpha,j)$ is labeled both by a quantum number $\alpha$ of the underlying QFT model and a copy number $j\in\{1,\ldots,n\}$. States will be denoted by $|\theta_1\cdots \theta_\ell\ket_{\mu_1,\ldots,\mu_\ell;n}$, and since the $n$ copies are independent, states are direct products, for instance
\beq
	|\theta_1 \theta_2\ket_{(\alpha_1,1),(\alpha_2,2);
	n}
	= |\theta_1\ket_{\alpha_1}\otimes |\theta_2\ket_{\alpha_2}
	\otimes|\vac\ket \otimes \cdots\otimes |\vac\ket
\eeq
where there are $n-2$ vacuum factors.

Branch-point twist fields, however, are not factorizable in this way and therefore their matrix elements are generally non-trivial. It was found \cite{entropy,next} that the two-particle form factors of the branch-point twist field have an extended physical strip. That is, the function
\beq
	F^{\mathcal{T}}_{\mu_1,\mu_2;n}(\theta_1-\theta_2):=\frc{
  {}_n\bra\vac|
  \mathcal{T}|\theta_1\theta_2\ket_{\mu_1,\mu_2;n}}{{}_n\bra\vac|\mathcal{T}|\vac\ket_n},
\eeq
defined as the analytic continuation of the matrix elements in terms of the variable $\theta=\theta_1-\theta_2$, is analytic in an {\em extended physical sheet}. Define
\beq
	F_{\alpha_1,\alpha_2;n}(\theta):= F^{\mathcal{T}}_{(\alpha_1,1),(\alpha_2,1);n}(\theta).
\eeq
This is analytic in the extended physical sheet ${\rm Im}(\theta)\in[0,2\pi n]$ except for possible {\rm bound state poles} and for {\em kinematic poles} with specific residues. It takes the values of in-state matrix elements on $(0,\infty)$ and $(2\pi i n -\infty,2\pi i n)$, and of out-state matrix elements on $(-\infty,0)$ and $(2\pi i n ,2\pi i n +\infty)$; and on the extended physical sheet the relation
\beq\label{2pii}
	F_{\alpha_2,\alpha_1;n}(2\pi i n-\theta)
	= F_{\alpha_1,\alpha_2;n}(\theta)
\eeq
holds. The function is analytic except for possible bound state poles on ${\rm Re}(\theta)=0$, ${\rm Im}(\theta)\in(0,\pi)\cup (2\pi  n-\pi,2\pi n)$, and except for kinematic poles at $\theta=i\pi$ and $\theta=2\pi i n-i\pi$ with residues $i$ and $-i$ respectively. Further, within the extended physical sheet, on strips of widths $2\pi i$ it takes the values of form factors with shifted copy numbers. Specifically, the relations
\beq\label{shifted}
	F^{\mathcal{T}}_{\mu_1,\mu_2;n}(\theta) =
	F_{\alpha_1,\alpha_2;n}(2\pi i \{j_1-j_2\}_n+\theta),
\eeq
where $\mu_1 = (\alpha_1,j_1)$ and $\mu_2=(\alpha_2,j_2)$, hold for all $j_1,j_2\in\{1,\ldots,n\}$; we define $\{j\}_n$ as the smallest non-negative integer in the set $\{j+nk:k\in\Z\}$ (that is, here we have $\{j_1-j_2\}_n=j_1-j_2$ if $j_1\geq j_2$ and $\{j_1-j_2\}_n = j_1-j_2+n$ if $j_1< j_2$). Note also the trivial  symmetry under cyclic permutation of the copies,
\beq\label{permsym}
	F^{\mathcal{T}}_{(\alpha_1,j),(\alpha_2,k);n}(\theta) =
	F^{\mathcal{T}}_{(\alpha_1,\{j+l\}_n),(\alpha_2,\{k+l\}_n);n}(\theta).
\eeq
Relations \eqref{2pii} and \eqref{shifted}, as well as the residue and boundary conditions, are consequences of the following equations, which hold in any massive QFT:
\beqa
	F^{\mathcal{T}}_{\mu_1,\mu_2;n}(\theta_1+2\pi i,\theta_2) &=& F^{\mathcal{T}}_{\mu_2,\mu_1+1;n}(\theta_2,\theta_1) \n
    F^{\mathcal{T}}_{\mu_1,\mu_2;n}(\theta_1,\theta_2) &=& F^{\mathcal{T}}_{\mu_2,\mu_1;n}(\theta_2,\theta_1)\quad (j_1\neq j_2) \n
    {\rm Res}_{\theta_1=\theta_2}
    F^{\mathcal{T}}_{\bar\mu_1,\mu_2;n}(\theta_1+\pi i,\theta_2) &=& 
    i\,\delta_{j_1,j_2}\delta_{\alpha_1,\alpha_2} \n
    {\rm Res}_{\theta_1=\theta_2}
    F^{\mathcal{T}}_{\bar\mu_1,\mu_2+1;n}(\theta_1+ \pi i,\theta_2) &=& 
    -i\,\delta_{j_1,j_2}\delta_{\alpha_1,\alpha_2} \
    \label{general}
\eeqa
where $\bar\mu$ denotes the anti-particle $(\bar\alpha,j)$ and $\mu+1$ denotes the copy-shifted particle $(\alpha,j+1)$.

{Additional poles are present whenever there are bound states. We will not need many details of these singularities, except for the following. Whenever a bound state of the particles $\alpha_1$ and $\alpha_2$ is present in the spectrum, with particle $\alpha_1$ coming from the left and $\alpha_2$ from the right, the function $F_{\alpha_1,\alpha_2;n}(\theta)$ possesses a pole at a purely imaginary rapidity. Such a pole is on the lower band of the extended physical sheet, at a position $\pi i y$ with $y\in(0,1)$, and has residue $i\gamma_{\alpha_1,\alpha_2;n}(y)\in\C$ proportional to the one-particle form factor of the branch-point twist field. Given $\alpha_1$ and $\alpha_2$, the set of values of $y$ corresponding to bound state poles on the lower band will be denoted $B_{\alpha_1,\alpha_2}$. By the relation \eqref{2pii}, similar poles occur on the upper band for all bound states corresponding to the particles $\alpha_1,\alpha_2$ being in the opposite order. These are at the positions $2\pi i n-\pi i y$ and have residues $-i\gamma_{\alpha_2,\alpha_1;n}(y)$ for all $y\in B_{\alpha_2,\alpha_1}$. Besides bound state and kinematic poles, there are no other singularities on the extended physical sheet. Note that many models posses parity invariance; in this case, $B_{\alpha_1,\alpha_2}=B_{\alpha_2,\alpha_1}$, and bound state poles occur in pairs with opposite residues, one on the lower band and one on the upper band. However we will not need to assume parity invariance.}

For convenience we will also use the notation $\gamma_{\alpha_1,\alpha_2;n}(1)=\delta_{\alpha_1,\b\alpha_2}$ for the kinematic residues at $y=1$, present both on the lower and upper band. We recall that the mass $m'$ of the bound state $y\in B_{\alpha_1,\alpha_2}$ of particles $\alpha_1$ and $\alpha_2$ of the same mass $m$ is given by $m' = 2m\cos(\pi y/2)$.

Note that the hermitian conjugate of the twist field is simply $\TT^\dag=\tilde{\TT}$, associated with the opposite cyclic copy permutation,
\beq\label{Tdagff}
	{}_n\bra\vac|\tilde{\TT}|\theta_1\theta_2\ket_{\mu_1,\mu_2;n} = {}_n\bra\vac|\mathcal{T}|\theta_1\theta_2\ket_{-\mu_1,-\mu_2;n}
\eeq
where $-\mu$ is interpreted as $(\alpha,n+1-j)$. Another relation that will be helpful is crossing symmetry, 
\beqa	{}_{\b\mu_2,\b\mu_1;n}\bra\theta_2\theta_1|\mathcal{T}|\vac\ket_{n} &=& 
 {}_n\bra\vac|\mathcal{T}|\theta_1\theta_2\ket_{\mu_1,\mu_2;n}.
\eeqa
Clearly, the left-hand side is the complex conjugate of a form factor of $\mathcal{T}^\dag$, and combining crossing symmetry with \eqref{Tdagff}, we obtain
\beq
	F^{\mathcal{T}}_{\mu_1,\mu_2;n}(\theta_1,\theta_2)
	=
	\lt(F^{\mathcal{T}}_{-\b\mu_2,-\b \mu_1;n}(\theta_2,\theta_1)\rt)^*.
\eeq
and in particular
\beq\label{stareq}
	F_{\alpha_1,\alpha_2;n}^*(\theta) = F_{\b\alpha_2,\b\alpha_1;n}(-\theta).
\eeq
Equation \eqref{stareq} implies that the set of bound state poles of $F_{\b\alpha_2,\b\alpha_1}(\theta)$ is the same as that of $F_{\alpha_1,\alpha_2}(\theta)$,
\beq\label{starB}
	B_{\alpha_1,\alpha_2} = B_{\b\alpha_2,\b\alpha_1}
\eeq
and that the residues are related to each other by complex conjugation:
\beq\label{stargamma}
	\gamma_{\alpha_1,\alpha_2}(y) = \gamma^*_{\b\alpha_2,\b\alpha_1}(y)\quad \forall\;y\in B_{\alpha_1,\alpha_2}.
\eeq
{In particular, $\gamma_{\alpha,\b\alpha}$ is real for all $\alpha$.}

We note that, due to relativistic invariance and invariance under permutation of copies, one-particle matrix elements $F_{(\alpha,j);n}^{\mathcal{T}}(\theta)$ are independent of $\theta$ and of the copy number $j$; we will denote them by $F_{\alpha;n}$. By similar arguments as those above, the following relation holds:
\beq\label{stareq1}
	\lt(F_{\alpha;n}\rt)^* = F_{\b\alpha;n}.
\eeq

Finally, in the calculation below we will need to perform an analytic continuation in the values of $n$. This analytic continuation is ambiguous, and usually, for twist field form factors in integrable models, one lifts the ambiguity by demanding that the large-$n$ behaviour be described by a series of increasing powers of $n^{-1}$. In the present case, since form factors are generic, this is hard to verify. We will instead assume that matrix elements of branch-point twist fields where all sheet numbers are equal to 1 have analytic expressions with correct behaviour at large $n$. We will assume that this analytic expression is such that the positions of poles in the lower part of the extended physical strip do not depend on $n$, and that of poles in the upper part depend on $n$ linearly, as $2\pi i n - \pi i y$ (so that in particular the set $B_{\alpha_1,\alpha_2}$ does not depend on $n$). We will also assume that these form factors vanish exponentially at large rapidities on the extended physical strip, and that they are not singular at $n=1/2$. All these properties will be verified explicitly in integrable models \cite{pre}.

\section{Universal scaling of the negativity}

The results (\ref{Zn3p}) and (\ref{e2}) indicate that a computation of the negativities $\mathcal{E}^{\perp}$ and $\mathcal{E}^{\dashv \, \vdash}$ will require the evaluation of the two point functions ${\,}_n\bra \text{vac}| \TT(0) \tilde{\TT}^2(r)| \text{vac}\ket_n $ and ${\,}_n\bra \text{vac}| \tilde{\TT}(0)\tilde{\TT}(r)| \text{vac}\ket_n$ for $n$ even. Such computations are routinely performed by inserting a decomposition of the identity in terms of asymptotic states
\beq\label{IDstates}
	{\bf 1} = \sum_{k=0}^\infty
	\sum_{\mu_1,\ldots,\mu_k}
	\int_{-\infty}^{\infty} \frc{d\theta_1\cdots d\theta_k}{(2\pi)^kk!}
	|\theta_1\cdots\theta_k\ket_{\mu_1,\ldots,\mu_k;n}
	\,
	{}_{\mu_1,\ldots,\mu_k;n}\bra\theta_1\cdots\theta_k|,
\eeq
between the two fields involved. In the expression \eqref{IDstates}, the integral over rapidities is unrestricted: this includes the basis of in-states and out-states, as well as generalized bases obtained by analytic continuation to other rapidity orders. The factor of $k!$ in the denominator accounts for this overcounting. Upon this insertion, the twist field correlators (and their logarithms) may be expressed as infinite sums over products of the form factors of the operators involved. It is well known that this sum naturally provides a large-distance expansion, where terms with higher numbers of particles decay faster as the distance $r$ increases.  Here we are interested in the leading (saturation) and next-to-leading order contributions to the negativity and this means that we will only consider contributions up to two-particle states. Let us now analyse each correlator in more detail. 

\subsection{Two adjacent regions}

Let us take the expression (\ref{Zn3p}) as our starting point. First, the behaviour at short distances can be obtained by using the zeroth order of conformal perturbation theory:
{\beq
	{}_n\bra\vac|\TT(0)\t\TT^2(r)|\vac\ket_n \sim C_n\,r^{-4\Delta_{n/2}} {\,}_n\bra\vac|\TT(0)|\vac\ket_n \qquad \mbox{
    ($n$ even)}
\eeq
where $C_n$ is the CFT the structure constant $C_{{\cal T}\t{\cal T}^2}^{\cal T}$; as shown in \cite{negMB} (see Eq. (90)), this equals $C_{{\cal T}{\cal T}}^{{\cal T}^2}$.} Here we used the fact that the conformal dimension of $\TT^2$ is given by \eqref{dimTT2} when $n$ is even. Using the definition (\ref{ZnEn}), analytically continuing to $n=1$ (from even values of $n$) and subtracting $\log(C_1)$ (through a re-definition of the non-universal short-distance cutoff $\varep$), this gives the first line of \eqref{main1}. On the other hand, in the limit where the distance $r$ is infinitely large, the two-point function factorizes and only the zero-particle term contributes. Thus (for $n$ even and using \eqref{T2}) the function factorizes into vacuum expectation values:
\beq
	{}_n\bra\vac|\mathcal{T}(0) \t{\mathcal{T}}^2(r)
	|\vac\ket_n \to {}_n\bra\vac|\mathcal{T}|\vac\ket_{n}\;
	{}_{\frac{n}{2}}\bra\vac|\mathcal{T}|\vac\ket_{\frac{n}{2}}^2.
\eeq
Since ${}_1\bra\vac|\mathcal{T}|\vac\ket_1=1$, employing the definition (\ref{ZnEn}) and again subtracting $\log(C_1)$ this implies that the negativity saturates to the value (\ref{U}).  The value of $\mathcal{E}_{\text{sat}}$ is model dependent and generally hard to evaluate. Expressions may be obtained for the vacuum expectation value of the twist fields in free models as reported in Appendix A, and we will provide a numerical analysis of the structure constants in a future work \cite{pre}.

Let us now consider the next-to-leading order correction to (\ref{U}). This can be obtained in all generality by analyzing the contribution of the two-particle terms in the spectral expansion of ${}_n\bra\vac|\mathcal{T}(r_1) \t{\mathcal{T}}^2(r_2)|\vac\ket_n$. This is because the contribution from one-particle form factors will always vanish when taking the limit $n\rightarrow 1$. Written in terms of form factors, and after performing one rapidity integral and expanding the logarithm function, one finds
\beq\label{exptineg1a}
	{\cal E}^\perp =-\frc c4 \log(m_1\varep)+ {\cal E}_{\rm sat} + \lim_{n\to1\atop n\;{\rm even}}
	\frc1{(2\pi)^2}\sum_{\mu_1,\mu_2}
	\int_{-\infty}^{\infty} d\theta\,
	F^{\mathcal{T}}_{\mu_1,\mu_2;n}(\theta)
	\lt(F^{\mathcal{T}^2}_{\mu_1,\mu_2;n}(\theta)\rt)^*
	K_0(M_{\alpha_1,\alpha_2}(\theta)\,r)+\ldots
\eeq
where here and below the ellipsis represents higher-particle contributions, and where
\beq
	M_{\alpha_1,\alpha_2}(\theta)
	=\sqrt{m_{\alpha_1}^2+m_{\alpha_2}^2 +
	2m_{\alpha_1}m_{\alpha_2}\cosh\theta} \label{man}
\eeq
is the square-root of Mandelstam's $s$-variable. We will show that this leading correction is exactly
\beq\label{sca1}\begin{aligned}
	{\cal E}^\perp = &\;-\frc c4 \log(m_1\varep)+ {\cal E}_{\rm sat}-\frc{2}{3\sqrt{3}\pi}\sum_{\alpha}K_0(\sqrt{3}m_\alpha r)\\ 
    &    -\,\frc1{3\pi}
	\sum_{\alpha}\,\sum_{ \substack{ y\in B_{\alpha,\b \alpha} \\ 
	\cap(1/2,1)}}
	\gamma_{\alpha,\b\alpha;\frc12}(y)
    \; \text{cosec}\left( \frac{\pi y + \pi}{3} \right)
    \, K_0\lt(
	2m_\alpha \cos\lt( \frc{2\pi y-\pi} 6\rt)r \rt)    
	\\
	&-\,\frc{1}{8\pi^2}
	\sum_{\alpha}
	\int_{-\infty}^{\infty} d\theta\,
	\cot \left(\frac{\pi -i\theta}{4}\right)
	\lt[
	F_{\alpha,\b\alpha;\frc 12}\left(\frac{\pi i +3\theta}{2}\right) 
	-
	F_{\alpha;\frc12}
	F_{\b\alpha;\frc12}
	\rt]
	K_0(2m_{\alpha}\cosh( \theta/2 )r)\\
    &+\,\ldots.
	\end{aligned}
\eeq
The higher-particle contributions are difficult to analyze in all generality, but an analysis in the massive free boson model \cite{pre} shows that they are $O(e^{-Zm_1r})$ for $Z>2$, and we expect this to hold in general with $m_1$ the smallest mass of the spectrum.

Equation \eqref{sca1} is the main result of this section, and is valid for any massive {unitary} one-dimensional QFT, integrable or not. The terms on the first line are the dominant terms, and give the second line of \eqref{main1}. Note the striking fact that, although the two-particle contribution to the twist field correlation function is of order $O(e^{-2m_1r})$ for any integer $n$, the analytic continuation in $n$ {\em changes the order of the leading correction to saturation} into a slower decay given by the above  $O(e^{-\sqrt{3}m_1r})$. We also remark that the discrete contributions from bound states (the second line in \eqref{sca1}) only come from the least massive ``half'' of the possible bound states between a particle $\alpha$ and its antiparticle, as $y\in B(\alpha,\b\alpha)\cap(1/2,1)$ implies masses in the range $(0,\sqrt 2 m_\alpha)$. These contributions give exponential decays with exponents between $\sqrt{3}m_\alpha$ and $2m_\alpha$ (ordered from the least massive to the most massive allowed bound state):
\beq
	\sqrt{3}m_\alpha <
	2m_\alpha \cos\lt( \frc{2\pi y-\pi} 6\rt)
	< 2m_\alpha.
\eeq
The integral contribution (the third line of \eqref{sca1}) is a sum over $\alpha$ of terms which are $O(e^{-2m_\alpha r})$. We note that it is real as
\[
	F_{\alpha,\b\alpha;\frc 12}\left(\frac{\pi i +3\theta}{2}\right)^*
	=
	F_{\alpha,\b\alpha;\frc 12}\left(\frac{\pi i -3\theta}{2}\right),\quad
	\lt(F_{\alpha;\frc12}\rt)^* = F_{\b \alpha;\frc12}
\]
by the crossing properties \eqref{stareq} and \eqref{stareq1}. Further, if every anti-particle is the particle itself, then one can symmetrize the cotangent factor by making the replacement
\[
	\cot\lt(\frc{\pi -i\theta}4\rt)\mapsto
	\frc12\, \frc{1-\tanh^2(\theta/4)}{1+\tanh^2(\theta/4)},
\]
as all other factors are symmetric, in particular thanks to property \eqref{2pii}.

Result \eqref{sca1} is shown as follows. Recall that the next-to-leading correction for the negativity comes from the expression (see \eqref{exptineg1a}) 
\beq
\label{intf}
	\frc1{(2\pi)^2}\sum_{\alpha_1,\alpha_2}
	\int_{-\infty}^{\infty} d\theta\,
	f_{\alpha_1,\alpha_2;n}(\theta)\,
	K_0(M_{\alpha_1,\alpha_2}(\theta)\,r)
\eeq
where the function $f_{\alpha_1,\alpha_2;n}(\theta)$ is given by
\beq
	f_{\alpha_1,\alpha_2;n}(\theta)
	:=\sum_{j_1,j_2=1}^n
	F^{\mathcal{T}}_{\mu_1,\mu_2;n}(\theta)
	\lt(F^{\mathcal{T}^2}_{\mu_1,\mu_2;n}(\theta)\rt)^*.
	\label{ff1}
\eeq
We will now consider $n$ even for the rest of this section.  The factorization property \eqref{T2} implies that the double sum naturally decomposes into sums over even and odd indices:
\beq
f_{\alpha_1,\alpha_2;n}(\theta)= f_{\alpha_1,\alpha_2;n}^{0,0}(\theta)+f_{\alpha_1,\alpha_2;n}^{0,1}(\theta)+f_{\alpha_1,\alpha_2;n}^{1,0}(\theta)+f_{\alpha_1,\alpha_2;n}^{1,1}(\theta)
\eeq
where $f_{\alpha_1,\alpha_2;n}^{k_1,k_2}(\theta)$ denotes the restriction of the sum in \eqref{ff1} to summation variables $j_1, j_2$ such that $j_1 \equiv k_1\; \text{mod} \; 2$ and $j_2 \equiv k_2\; \text{mod} \;2$.   
%
%
Let us first concentrate on $f_{\alpha_1,\alpha_2;n}^{0,0}(\theta)$ and $f_{\alpha_1,\alpha_2;n}^{1,1}(\theta)$. One observes that these are identical using the relations \eqref{T2}, \eqref{permsym} and \eqref{shifted}:
\beq \begin{split}
	& \sum_{j_1,j_2\;{\rm even}}
	F^{\mathcal{T}}_{\mu_1,\mu_2;n}(\theta)
	\lt(F^{\mathcal{T}}_{\frc{\mu_1}2,\frc{\mu_2}2;\frc n2}(\theta)\rt)^* =
	\sum_{j_1,j_2\;{\rm odd}}
	F^{\mathcal{T}}_{\mu_1,\mu_2;n}(\theta)
	\lt(F^{\mathcal{T}}_{\frc{\mu_1+1}2,\frc{\mu_2+1}2;\frc n2}(\theta)\rt)^* \\ & \qquad\qquad\qquad\qquad= \frc n2 \sum_{j=0}^{\frac{n}{2}-1}
	F_{\alpha_1,\alpha_2;n}(4\pi i j+\theta)
	\lt(F_{\alpha_1,\alpha_2;\frc n2}(2\pi i j+\theta)\rt)^*
	\end{split}
    \eeq
where the algebraic operations on $\mu=(\alpha,j)$ are interpreted as operations on the copy number $j$. Adding both and using crossing symmetry \eqref{stareq}, we obtain
\beq\label{f1gen}
	f_{\alpha_1,\alpha_2;n}^{0,0}(\theta)+f_{\alpha_1,\alpha_2;n}^{1,1}(\theta) = n
    \sum_{j=0}^{n/2-1}
    F_{\alpha_1, \alpha_2 ; n}(4\pi i j+\theta) F_{\bar{\alpha}_2, \bar{\alpha}_1 ;  n/2}(2\pi i j-\theta). 
\eeq

In order to analytically continue the sum in $n$, we separate the term with $j=0$. This term is a product of form factors whose arguments stay within the extended physical sheet for all $n>1$, hence we may directly analytically continue the resulting integral. Clearly $F_{\alpha_1, \alpha_2 ; 1}$ is zero as it is the two-particle form factor of the identity field, and no pole of $F_{\alpha_1, \alpha_2 ; n}$ cross the integration contour (the real line) as the limit $n\to1$ is taken. The function $F_{\bar{\alpha}_2, \bar{\alpha}_1 ;  1/2}$ is non-zero, and the kinematic pole of $F_{\bar{\alpha}_2, \bar{\alpha}_1 ;  n/2}$ in the upper band is brought to the real line from above upon taking $n\to1^+$, as $2\pi i (n/2)-i\pi=0$ for $n=1$. Under the integration in \eqref{intf}, the analytic continuation is that where the contour avoids this pole from below. The function $F_{\bar{\alpha}_2, \bar{\alpha}_1 ;  1/2}$ is otherwise expected to be finite (which we have explicitly checked in free models and in the integrable sinh-Gordon model). Hence, with $F_{\alpha_1, \alpha_2 ; 1}=0$, the result of the integral is zero at $n=1$, and we conclude that this term does not contribute to the negativity.

The sum from $j=1$ to $j=n/2-1$ is evaluated by a standard use of Cauchy's theorem.  Let $s(z):=F_{\alpha_1, \alpha_2 ; n}(4\pi i z+\theta) F_{\bar{\alpha}_2, \bar{\alpha}_1 ;  n/2}(2\pi i z-\theta)$, which is regular at the points $z\in\{1,2,\ldots,n/2-1\}$. Cauchy's theorem gives the following expression for the sum of $s(j)$:
\beq\label{sum}
	\sum_{j=1}^{\frac{n}{2}-1} s(j) = \frc1{2 i} \oint_{\cal C} dz\, \cot( \pi z) \,s(z) - \sum_{i}
	\pi\cot(\pi z_i^\star )\, \text{Res}[  s(z) \; ; \, z=z_i^\star]
\eeq
where ${\cal C}$ is a counter-clockwise coutour that surrounds the points in $\{1,2,\ldots,n/2-1\}$ but not those in $\Z \setminus \{1,2,\ldots,n/2-1\}$, and where $\{z_1^\star,z_2^\star, \ldots\}$ is the set of poles of $s(z)$ surrounded by ${\cal C}$. For the present purpose, $\cal C$ is chosen to be composed of two vertical lines on ${\rm Re}(z)=0^+$ and ${\rm Re}(z)=n/2-0^+$. The function $s(z)$ vanishes exponentially at $z\to\pm i\infty$, so we omit the pieces at $\pm i\infty$ that close the contour. On the vertical line ${\rm Re}(z)=0$ the function $s(z)$ is a product of form factors with real arguments, and using \eqref{2pii} the same observation holds for ${\rm Re}(z) = n/2$. Hence for the reasons explained above, the analytic continuation to $n=1$ of these products vanishes, whereby for evaluating the negativity we may neglect the integral part in \eqref{sum}.

Note that the integration contour stays within the extended physical sheet, so that we can use the known analytic properties there. Recall that $B_{\alpha_1,\alpha_2}$ characterizes the set of bound-state singularities of $F_{\alpha_1,\alpha_2;n/2}(\theta)$ (if any) {in the lower band of the physical strip, and that $B_{\alpha_2,\alpha_1}$ characterizes that in the upper band}, and recall the relation \eqref{starB}. By the principles stated in section \ref{sectscatt}, the function $s(z)$ has poles at the points $z^\star_i$ with residues $r^\star_i$ as follows:
{\beqa
	z^\star_1 = \frc{y}4 - \frc{\theta}{4\pi i} &:&
		r^\star_1 = \frc{\gamma_{\alpha_1, \alpha_2 ; n }(y)}{4\pi}\, F_{ \bar{\alpha}_2, \bar{\alpha}_1  ; \frc n2}\lt(\frc{\pi i y}2-\frc{3\theta}2\rt) \n
	z^\star_2 = \frc{y}2 + \frc\theta{2\pi i} &:&
		r^\star_2 = \frc{\gamma_{\b\alpha_2, \b\alpha_1 ; n/2 }(y)}{2\pi}\,F_{\alpha_1, \alpha_2 ; n}\lt(2\pi iy + 3\theta\rt) \no
\eeqa
for each $y\in B_{\alpha_1, \alpha_2}$, and
\beqa
        z^\star_3 = \frc{n}2 - \frc{y}4 -\frc\theta{4\pi i} &:&
		r^\star_3 = -\frc{\gamma_{\alpha_2, \alpha_1 ; n}(y)}{4\pi}\, F_{\bar{\alpha}_2, \bar{\alpha}_1 ; \frc n2}\lt(\pi i n-\frc{ \pi i y}2-\frc{3\theta}2\rt) \n
	z^\star_4 = \frc n2- \frc{y}2 + \frc\theta{2\pi i} &:&
		r^\star_4 = -\frc{\gamma_{\b\alpha_1, \b\alpha_2 ; n/2 }(y)}{2\pi}\,F_{\alpha_1, \alpha_2 ; n}\lt(2\pi i n-2\pi i y + 3\theta\rt).\no
\eeqa
for each $y\in B_{\alpha_2, \alpha_1}$.} In addition, the function $s(z)$ has kinematic poles if and only if $\alpha_1=\b\alpha_2$, four poles given by the above values of $z_i^\star$ and $r_i^\star$ with $y=1$ and $\gamma_{\alpha,\b\alpha;n}(1)=1$. Recall that each bound state residue is proportional to a one-particle form factor of the branch-point twist field (independent of the rapidity and of the copy number), and note that $\gamma_{\alpha_1, \alpha_2 ; 1 }(y)=0$ for evey $y\in B_{\alpha_1,\alpha_2}$, and that $\gamma_{\alpha_1, \alpha_2 ; 1/2 }(y)$ is expected to be finite.  The contribution of the residues to $f^{0,0}_{\alpha_1,\alpha_2;n}(\theta) + f^{1,1}_{\alpha_1,\alpha_2;n}(\theta)$ sums up to 
{\begin{equation} \label{polegen} \begin{split}
-\frac{n}{4}\sum_{y\in B_{\alpha_1,\alpha_2}\cup\{ 1\}} & \Big[
\gamma_{\alpha_1,\alpha_2;n}(y) \cot \left(\frac{\pi y + i\theta}{4}\right)F_{ \bar{\alpha}_2, \bar{\alpha}_1 ; \frc n2}\left(\frac{\pi i y-3\theta}{2}\right) \\
&
+ 2\gamma_{\b\alpha_2,\b\alpha_1;\frc n2}(y) \cot \left(\frac{\pi y - i\theta}{2}\right)F_{ {\alpha}_1, {\alpha}_2 ;  n}( 2 \pi i y+3\theta)\Big]\\
-\frac{n}{4}\sum_{y\in B_{\alpha_2,\alpha_1}\cup\{ 1\}} & \Big[
\gamma_{\alpha_2,\alpha_1;n}(y) \cot \left(\frac{\pi y - i\theta}{4}\right)F_{ \bar{\alpha}_1, \bar{\alpha}_2 ; \frc n2}\left(\frac{\pi i y+3\theta}{2}\right) \\
&
+ 
2\gamma_{\b\alpha_1,\b\alpha_2;\frc n2}(y) \cot \left(\frac{\pi y + i\theta}{2}\right)F_{ {\alpha}_2, {\alpha}_1 ;  n}( 2 \pi i y-3\theta)
\Big]
\end{split}
\end{equation}}
\noindent
where we used the fact that $n$ is even in order to simplify some of the trigonometric factors. Note that the sum on $y$ runs over the bound state poles $y\in B_{\alpha_1,\alpha_2}$ and $y\in B_{\alpha_2,\alpha_1}$, and the kinematic pole $y=1$ (where all $\gamma$'s specialize to 1).

The contribution of \eqref{polegen} to the integral \eqref{intf} is then
{\beq\begin{split}
-\frac{n}{ 8\pi^2} \sum_{ \alpha_1, \alpha_2}\;\sum_{y\in B_{\alpha_1, \alpha_2} \cup\{1\} } &
\int_{-\infty}^{\infty} d\theta\, K_0(M_{\alpha_1, \alpha_2}(\theta) r)\,\times \\ & \times  \Bigl[ 
\cot\left(
\frac{\pi y-i\theta}{4} \right)
\gamma_{\alpha_1,\alpha_2;n}(y) 
F_{\b{\alpha}_2, \b{\alpha}_1;\frac n2} \left(\frac{\pi i y + 3 \theta}{2} \right)
\\
& + 2 
\cot\left(  \frac{\pi y-i\theta}{2} \right)
\gamma_{\b\alpha_2,\b\alpha_1;\frc n2}(y)
F_{{\alpha}_1, {\alpha}_2; n} (2\pi i y + 3 \theta )
\Bigr]
\end{split}
\label{intf2}
\eeq}
where we have used the fact that the integrand is invariant under $\theta \rightarrow -\theta$, {and we have use a change of summation indices $\alpha_1,\alpha_2\mapsto \alpha_2,\alpha_1$ in order to have a single $y$ sum}. The $n$-dependence is now explicit and analytic, and since the arguments of every form factor involved stay within their extended physical sheet for all values of $n>1$, we may directly take the limit $n\to1$ in \eqref{intf2}.

Substituting $n=1$ in the integrand of \eqref{intf2}, the only nonzero terms are those corresponding to the kinematic residue contribution in the first line inside the square brackets, with $\alpha_1=\b\alpha_2$ and $y=\gamma_n=1$ (this is because the form factors and bound-state $\gamma$'s at $n=1$ are zero). This gives
\beq\label{c0}
	-\frc{1}{8\pi^2} \sum_{\alpha} 
	\int_{-\infty}^{\infty} d\theta \, K_0(2m_\alpha \cosh(\theta/2) r ) \,
	\cot \left(\frac{\pi -i\theta}{4}\right)
    F_{\alpha,\b\alpha;\frc 12}
    \left(\frac{\pi i +3\theta}{2}\right).
\eeq
The limit $n\to1$ of the analytic function \eqref{intf2}, taken from large enough values of $n$, receives additional contributions: the residues of the poles that cross the integration contour as the limit is taken.

Consider the first line inside the square brackets in \eqref{intf2}. Because of the factor $\gamma_{\alpha_1,\alpha_2;n}$, the only nonzero contributions as $n\to1$ come from the kinematic pole: we restrict the particle-species sum to $\alpha_1=\b\alpha_2=:\alpha$, and the pole sum to the single term with $y=1$ (at which the $\gamma$'s are equal to 1). In this case, {the form factor in the parentheses specializes to} $F_{\alpha,\b\alpha;n/2}((\pi i +3\theta)/2)$, which has argument on the line with imaginary part $\pi /2$. Recall that its poles (kinematic and bound-state) on the upper band of the extended physical sheet are at $2\pi i (n/2)-\pi i y'$ for $y'\in (0,1]$ (the bound-state values $y'\in B_{\b\alpha,\alpha}$ and the kinematic value $y'=1$). As $n\to1$, the form factor goes from a high-copy-number to copy number $1/2$. As $n$ reaches 1, the poles on the upper band have reached the segment $[0,\pi i)$. The set of these poles in $[0,\pi i/2)$ have crossed the integration line; these are for $y'\in(1/2,1]$. The residues taken have $\theta$-value, in the limit $n\to1$, given by
\[
	\frc{\pi i + 3\theta}2 = \pi i(1-y')\;\Rightarrow\;
	\theta = \frc{\pi i(1-2y')}3,
\]
and the part of the residue coming from the form factor itself, $F_{\alpha,\b\alpha;n/2}((\pi i +3\theta)/2)$ is, in the limit $n\to1$, given by $-2i\gamma_{\b\alpha,\alpha;1/2}(y')/3$. Evaluating the corresponding Cauchy contributions on the first line inside the square brackets in \eqref{intf2}, with the factor $2\pi i$ as the Cauchy contour around each pole is counter-clockwise, we find
\beq\label{c1}
	-\frc{1}{2\sqrt{3}\pi}\sum_{\alpha}K_0(\sqrt{3}m_\alpha r)
	-\frc1{6\pi}
	\sum_{\alpha}\,\sum_{y\in B_{\alpha,\b \alpha}\atop \cap(1/2,1)}
	\gamma_{\alpha,\b\alpha;\frc12}(y) \tan\lt(\frc{\pi y + \pi}{6}\rt)
	\, K_0\lt(
	2m_\alpha \cos\lt( \frc{2\pi y-\pi} 6\rt)r \rt)
\eeq
where the first term is for the kinematic pole ($y=1$) and the second for the allowed bound state poles as described {(under the change of summation index $\alpha\mapsto\b\alpha$)}.

Now consider the second line inside the square brackets in \eqref{intf2}. There, the {form factor $F_{{\alpha}_1, {\alpha}_2; n} (2\pi i y + 3 \theta )$ has argument} on the line with imaginary part $2\pi y$. Recall that its poles on the upper band of the extended physical sheet are at $2\pi i n-\pi i y'$ for $y'\in (0,1]$ (bound-state and kinematic values). The subset of these poles satisfying $2\pi - \pi y'< 2\pi y$ cross the integration line upon taking the limit $n\to1$. Note that for $y\leq 1/2$, none of the poles cross. The bound state poles have residues $\gamma_{\alpha_2,\alpha_1;n}(y')$, which vanish at $n=1$. Hence, the only pole that will give a contribution is the kinematic pole, at $y'=1$, which crosses the integration contour if and only if $y>1/2$. Hence, again this constrains the particle-species sum in \eqref{intf2} to $\alpha_1=\b\alpha_2=:\alpha$ and the sum over bound state / kinematic positions to $y>1/2$. In the limit $n\to1$, the $\theta$-value of the crossed kinematic pole is
\[
	2\pi i y + 3\theta = \pi i \;\Rightarrow\;
	\theta = \frc{\pi i (1-2y)}3,
\]
and the part of the residue coming from the factor $F_{{\alpha}, \b\alpha; n} (2\pi i y + 3 \theta )$ is $-i/3$. Evaluating the corresponding Cauchy contributions on the second line inside the square brackets in \eqref{intf2}, again with the factor $2\pi i$ as the Cauchy contour around each pole is counter-clockwise, we find
\beq\label{c2}
	-\frc{1}{6\sqrt{3}\pi}\sum_{\alpha}K_0(\sqrt{3}m_\alpha r) \\
	- \frc1{6\pi} \sum_{\alpha}\,\sum_{y\in B_{\alpha,\b\alpha}\atop
	\cap(1/2,1)}
	\gamma_{\alpha,\b\alpha;\frc12}(y) \,\cot\lt(\frc{\pi y + \pi}6\rt)\,
	K_0\lt(
	2m_\alpha \cos\lt( \frc{2\pi y-\pi} 6\rt)\,r \rt)
\eeq
where the first term is for the kinematic pole $y=1$ and the second for the allowed bound state poles.

Finally, we analyze $f_{\alpha_1,\alpha_2;n}^{0,1}(\theta)$ and $f_{\alpha_1,\alpha_2;n}^{1,0}(\theta)$, the sums over mixed (even-odd) indices. In this case, the form factors of $\mathcal{T}^2$ factorize into products of one-particle form factors (which are independent of the rapidity and of the copy number):
\beq \begin{split}
	f_{\alpha_1,\alpha_2;n}^{0,1}(\theta)
    =
    f_{\alpha_2,\alpha_1;n}^{1,0}(-\theta)
	&= \sum_{j_1\;{\rm even}\atop \,j_2\;{\rm odd}}
	F^{\mathcal{T}}_{\mu_1,\mu_2;n}(\theta)
	F_{\alpha_1;\frc n2}^*
	F_{\alpha_2;\frc n2}^* \\
	&= \frc n2
	\sum_{j=0}^{\frac{n}{2}-1}
	F_{\alpha_1,\alpha_2;n}(2\pi i(2j+1)+\theta)
	F_{\b\alpha_1;\frc n2}
	F_{\b\alpha_2;\frc n2} 
\end{split}
\eeq
where we used \eqref{stareq1} and the second equation of \eqref{general}. We again use formula \eqref{sum} in order to evaluate the sum, with $s(z) = F_{\alpha_1,\alpha_2;n}(2\pi i(2z+1)+\theta)$ and choosing a counterclockwise contour formed by the two vertical lines ${\rm Re}(z) = -1/2$ and ${\rm Re}(z) = n/2-1/2$ (these are the lines that bound the extended physical strip). The result is
{\beq\begin{split}
	f_{\alpha_1,\alpha_2;n}^{0,1}(\theta)
	&\,=\,
	\frc n8 \,F_{\b\alpha_1;\frc n2}
	F_{\b\alpha_2;\frc n2}\,\times \\ &\times
	\lt(
    \sum_{y\in B_{\alpha_1,\alpha_2}\cup\{1\}} 
    \gamma_{\alpha_1,\alpha_2;n}(y)
    \tan\lt(\frc{\pi y + i\theta}4\rt)
    +
    \sum_{y\in B_{\alpha_2,\alpha_1}\cup\{1\}} 
    \gamma_{\alpha_2,\alpha_1;n}(y)
    \tan\lt(\frc{\pi y - i\theta}4\rt)
    \rt)
    \end{split}
\eeq}
where the sum is over the bound states $y\in B_{\alpha_1,\alpha_2}$ (for the lower band of the extended physical strip) and $y\in B_{\alpha_2,\alpha_1}$ (for the upper band), and the kinematic pole $y=1$ (which contributes if and only if $\alpha_1=\b\alpha_2$). Putting this inside the integral in \eqref{intf} and taking the limit $n\to1$, only the term in the sum corresponding to the kinematic pole contributes, as all other terms gives $\gamma_{\alpha_1,\alpha_2;1}=0$. The result, including both contributions from integrating $f_{\alpha_1,\alpha_2;n}^{0,1}(\theta)$ and $f_{\alpha_1,\alpha_2;n}^{1,0}(\theta)$ (which are the same), can be written as
\beq\label{c3}
	\frc{1}{8\pi^2}
	\sum_{\alpha}
	\int_{-\infty}^{\infty} d\theta\,
	\cot \left(\frac{\pi -i\theta}{4}\right)
	F_{\alpha;\frc12}
	F_{\b\alpha;\frc12}\,
	K_0\lt(2m_{\alpha}\cosh\lt(\frc \theta2\rt)r\rt).
\eeq
Combining \eqref{c0}, \eqref{c1}, \eqref{c2} and \eqref{c3}, we find \eqref{sca1}.

\subsection{Two semi-infinite non-adjacent regions}

Consider now the function (\ref{e2}). The short distance behaviour is obtained by zeroth-order conformal perturbation theory,
\beq
	{}_n\bra\vac|\t{\mathcal{T}}(0) \tilde{\mathcal{T}}(r)
	|\vac\ket_n
    \sim C_n \,r^{-4\Delta_n + 4\Delta_{n/2}}
    {}_n\bra\vac|\tilde{\mathcal{T}}^2|\vac\ket_n
\eeq
where we used \eqref{dimTT2} for $n$ even, and conformal normalizations of the fields. Again $C_n$ is the CFT structure constant $C_{{\cal T}{\cal T}}^{{\cal T}^2}$. This gives rise to the first line of \eqref{sca2}.

On the other hand, the 
leading behaviour of the logarithmic negativity at large distances is dictated by the factorization
\beq 
 {}_n\bra\vac|\t{\mathcal{T}}(0) \tilde{\mathcal{T}}(r)
	|\vac\ket_n \to {}_n\bra\vac|\tilde{\mathcal{T}}|\vac\ket_{n}^2.
\eeq
Thus, the negativity will decrease as $r \rightarrow \infty$ to a constant value which in this case is simply $2 \log({}_1\bra\vac|\mathcal{T} |\vac\ket_1)=0$ for all theories, in agreement with physical intuition (there should be no entanglement between infinitely separated regions). The entanglement build-up as regions get closer is obtained again by means of a form factor expansion which takes the following form\beq
	{\cal E}^{\dashv \, \vdash} = \lim_{n\to1\atop n\;{\rm even}}
	\frc1{4\pi^2}\sum_{\mu_1,\mu_2}
	\int_{-\infty}^\infty d\theta\,
	F^{\tilde{\TT}}_{\mu_1,\mu_2;n}(\theta)
	\lt(F^{{\TT}}_{\mu_1,\mu_2;n}(\theta)\rt)^*
	K_0(M_{\alpha_1,\alpha_2}(\theta)\, r)
    +\ldots, \label{chula}
\eeq
where ellipsis are higher-particle contributions, and
where $M_{\alpha_1,\alpha_2}(\theta)$  is as in (\ref{man}); the one-particle form factor contribution is once again not written explicitly as it is vanishing for $n\rightarrow 1$.
We will now show that this leading correction is exactly
\beq\label{sca2p}
	{\cal E}^{\dashv \, \vdash} =  \frac{1}{2\pi^2} \sum_{\alpha}(m_\alpha r) \left[m_\alpha r K_0(m_\alpha r)^2+K_0(m_\alpha r)K_1(m_\alpha r)-m_\alpha r K_1(m_\alpha r)^2\right] + \ldots
	\eeq
giving the second line of \eqref{sca2}, where the higher-particle contributions are expected to be of the order of the 4-particle contributions to the two-point function.

Let us consider the sum
\beq 
g_{\alpha_1,\alpha_2;n}(\theta)=\sum_{j_1,j_2=1}^n F^{\tilde{\TT}}_{\mu_1,\mu_2;n}(\theta)
	\lt(F^{{\TT}}_{\mu_1,\mu_2;n}(\theta)\rt)^*.
\eeq 
In this case we are summing over all indices (even or odd) so this is a much easier sum to perform than (\ref{ff1}). We can simply exploit the properties of form factors \eqref{shifted}, \eqref{permsym} and \eqref{stareq}, as well as \eqref{Tdagff}, in order to write
\beq
g_{\alpha_1,\alpha_2;n}(\theta)=n\sum_{j=0}^{n-1} F_{\alpha_1,\alpha_2;n}(2\pi i j+\theta)\,
F_{\b\alpha_1,\b\alpha_2;n}(2\pi i j+\theta).
\eeq 
The sum above may be computed in much the same way as the sum (\ref{f1gen}), namely by exploiting Cauchy's theorem exactly as in (\ref{sum}) with $n/2$ replaced by $n$ and $s(z)=F_{\alpha_1,\alpha_2;n}(2\pi i z+\theta)
F_{\b\alpha_1,\b\alpha_2;n}(2\pi i z+\theta)$.
The integration contour ${\cal C}$ is again taken to be composed of two vertical lines on the boundaries of the extended physical sheet. As before, the part of the integration on the vertical lines vanish as $n\to1$, hence can be neglected for the evaluation of the negativity. Therefore, we concentrate on computing the residues of the poles falling within the integration contour. In general we will again have both kinematic and bound state poles. The kinematic poles are  $z_1 = \frc{1}2 - \frc{\theta}{2\pi i}$ and $z_2 =n- \frc{1}2 - \frc{\theta}{2\pi i}$ and they occur when $\alpha_1=\bar{\alpha}_2$. The bound state poles  are  $\frc{y}2 - \frc{\theta}{2\pi i}$ and $n- \frc{y}2 - \frc{\theta}{2\pi i}$ and they occur for all {$y\in B_{\alpha_1,\alpha_2}\cup B_{\alpha_2,\alpha_1}$}. The bound state poles will necessarily give contributions proportional to the residues $\gamma_{\alpha_1,\alpha_2;n}(y)$ or $\gamma_{\alpha_2,\alpha_1;n}(y)$ {(or their complex conjugates)}. As these vanish as $n\to1$, we may neglect their contribution for the evaluation of the negativity. Hence we restrict to $\alpha_1=\bar{\alpha}_2=:\alpha$ and only consider the kinematic poles.

The kinematic poles of the form factors give rise to double poles in the function $F_{\alpha,\b\alpha;n}(2\pi i j+\theta)
F_{\b\alpha,\alpha;n}(2\pi i j+\theta)$, so the computation of the corresponding residues is a bit more involved. 
Let us consider the form factors involved. Near the kinematic poles,  they have the form
\beq
	F_{\alpha,\b\alpha;n}(\theta) \sim \frc{i}{\theta-i\pi} + \kappa_\alpha,\quad
    F_{\alpha,\b\alpha;n}(\theta) \sim \frc{-i}{\theta - 2\pi i n + i\pi} + \kappa_{\b\alpha}
\eeq
for some constants $\kappa_\alpha$, where we used the property \eqref{2pii}, $F_{\alpha,\b\alpha;n}(2\pi i n -\theta) = F_{\b\alpha,\alpha;n}(\theta)$, in order to relate the constants on both poles. Let us compute the corresponding residues of the function $s(z) \cot \pi z$.  Around the pole $z_1$ we have the leading divergency
\beq
	\sim \lt(\frc{i}{2\pi i (z-z_1)} + \kappa_\alpha\rt)
    \lt(\frc{i}{2\pi i (z-z_1)} + \kappa_{\b\alpha}\rt)\cot \pi z
    = \ldots + \frc{(\kappa_\alpha+\kappa_{\b\alpha})\cot \pi z_1}{2\pi(z-z_1)} -
    \frc{{\rm cosec}^2 \pi z_1}{4\pi (z-z_1)} + \ldots
\eeq
and $2\pi i$ times the residue then gives
\beq
	i(\kappa_\alpha+\kappa_{\b\alpha}) \cot\pi z_1 - \frc{i}{2\cosh^2(\theta/2)}.
\eeq
Similarly, around $z_2$, we find
\beq
	\sim \lt(\frc{-i}{2\pi i (z-z_2)} + \kappa_{\b\alpha}\rt)
    \lt(\frc{-i}{2\pi i (z-z_2)} + \kappa_\alpha\rt)
    \cot \pi z
    = \ldots - \frc{(\kappa_\alpha+\kappa_{\b\alpha})\cot \pi z_2}{2\pi(z-z_2)} -
    \frc{{\rm cosec}^2 \pi z_2}{4\pi (z-z_2)} + \ldots
\eeq
and $2\pi i$ times the residue then gives
\beq
	-i(\kappa_\alpha+\kappa_{\b\alpha})\cot\pi z_2 - \frc{i}{2\cosh^2(\theta/2)}.
\eeq
Adding these two contributions,  using $\cot\pi z_1 = \cot\pi z_2 = -\tan(\theta/2i)$ and finally dividing by $-2i$ and multiply by $n$ we get the exact contribution to the sum, 
\beq
	\frac{n}{2}{\rm sech}^2 \frac{\theta}{2}, \label{con1}
\eeq
from the kinematic poles. Inserting the contribution (\ref{con1})  into the integral (\ref{chula}) and taking the limit $n\rightarrow 1$ we obtain the following leading correction to the negativity:
\beq 
{\cal E}^{\dashv \, \vdash} = \frc1{4\pi^2} \sum_{\alpha}\int_{0}^\infty d\theta \, {\rm sech}^2 \frac{\theta}{2}\,
	K_0(2 m_\alpha \cosh\left(\frac{\theta}{2}\right) r).\eeq 
This integral may be computed exactly giving the universal result (\ref{sca2p}).

\section{Conclusions and outlook}
In this paper we have considered the logarithmic negativity, a measure of bipartite entanglement, in a general 1+1-dimensional massive quantum field theory, not necessarily integrable. We have considered two particularly interesting limiting cases. Firstly, we have computed the negativity between a finite region of length $r$ and an adjacent semi-infinite region. Secondly we have computed the negativity between two semi-infinite regions separated by a distance $r$. The main results of our analysis can be summarized as follows:

\begin{itemize}
\item[1)] Both limits of the negativity tend to a constant for $r\rightarrow \infty$. In the latter case, this constant is zero, which agrees with physical intuition according to which there is no entanglement between infinitely separated regions. For adjacent regions, the saturation constant is theory-dependent, requiring the knowledge of the expectation values of twist fields and of the twist field three point coupling $C_{\TT \TT}^{\TT^2}$. We have provided explicit formulae for the contribution to saturation of these expectation values in free Majorana and free Boson theories (see Appendix A). However, general formulae for the three point coupling involved are not yet known, even for free models. 

\item[2)] The leading corrections to saturation are (in both limits) exponential decays in $r$ (described by modified Bessel functions) that are solely controlled by the mass spectrum of the model, independently of its scattering matrix. This implies that, like the entanglement entropy, the logarithmic negativity displays a very high level of universality, allowing one to extract information about the mass spectrum.

\item[3)] By studying sub-leading terms we have shown that, unlike the entanglement entropy \cite{entropy,next}, a large-$r$ analysis of the negativity of adjacent regions allows for the detection of bound states.
\end{itemize}

There are still many additional features of the negativity we would like to explore for 1+1 dimensional QFT. From the numerical point of view, it would be very interesting to test our present predictions numerically by evaluating corrections to the negativity of gapped quantum spin chains. It should also be possible to exploit the form factor approach in order to obtain numerical estimates of the CFT three point coupling $C_{\TT \TT}^{\TT^2}$ in free theories. From the point of view of analytical computations, it would be natural and interesting to analyze further sub-leading corrections to the limiting cases of the negativity considered here. Preliminary results \cite{pre} suggest that some of these corrections may still be universal. For the case of two semi-infinite regions separated by a distance $r$ we expect that a full description of all corrections may be accessible for free theories by performing an analysis reminiscent of the study carried out in \cite{nexttonext}.

A more general study of the negativity will involve the evaluation of more general matrix elements of the form ${\,}_n\bra\theta_1, \cdots, \theta_k| \TT| \beta_p \cdots \beta_1\ket_n$ with all the subtleties that are well known for this kind of problem and the additional complication of considering twist fields. We expect to report progress in this direction for integrable models in the near future \cite{pre}. 

\appendix 

\section{Formulae for the free Majorana and free Boson theories}

The general results (\ref{sca1}) and (\ref{sca2p}) can be easily specialized to the free Majorana (FM) and free Boson (FB) theories (with mass $m$). In the case of (\ref{sca2p}) this is immediate whereas (\ref{sca1}), as we will see, exhibits some model dependence in the subleading corrections. The FM and FB theories are very simple, consisting of a single particle spectrum with no bound states and two-particle scattering matrices given by $S(\theta)=-1$ and $S(\theta)=1$, respectively. Due to internal symmetries, branch point twist field form factors are only non-vanishing for even particle numbers (even for general $n$). In particular, the two-particle form factors associated to two particles on the same copy (normalized by the vacuum expectation value) are given by:
\beq 
F_n^{\text{FM}}(\theta)=\frac{\sin\frac{\pi}{n}}{2n \sinh\left(\frac{i\pi+\theta}{2n}\right)\sinh\left( \frac{i\pi-\theta}{2n}\right)} \frac{\sinh\frac{\theta}{2n}}{\sinh\frac{i\pi}{2n}}\quad \text{and} \quad 
F_n^{\text{FB}}(\theta)=\frac{\sin\frac{\pi}{n}}{2n \sinh\left(\frac{i\pi+\theta}{n}\right)\sinh\left( \frac{i\pi-\theta}{2n} \right)},
\eeq 
and all higher particle form factors may be obtained by employing Wick's theorem. For the FM theory these form factors can be written in terms of a Pfaffian determinant \cite{nexttonext}. The same formula may be used for the FB theory by simply replacing all terms with negative sign by terms with a positive sign. 

The values of $\mathcal{E}_{\text{sat}}$ and $\mathcal{E}_{\text{shift}}$ given by \eqref{U} and \eqref{Ud} are model dependent and generally hard to evaluate. Although we do not have expressions for the CFT structure constant involved, expressions may be obtained for the twist field vacuum expectation values in free models, thus giving $\frc{\mathcal{E}_{\rm shift}+\mathcal{E}_{\rm sat}}2$ \eqref{vacc}. For the FM theory the expectation values of twist fields are known \cite{entropy} for generic values of $n$ and the following value can easily be obtained:
\beqa 
	\frc{\mathcal{E}_{\rm shift}^{\rm FM}
    +\mathcal{E}_{\rm sat}^{\rm FM}}2
    & =&  \frac{1}{8} \log 2+ \int_0^\infty \frac{dt}{2t}\left[ \frac{e^{-2t}}{4}+\frac{1-\cosh t}{\sinh t \sinh (2t)}\right]\nonumber\\
    &=&-\frac{1}{8}+\frac{\log 2}{3}+ \frac{3 \log A}{2}-\frac{1}{2}\log\left(\frac{3\Gamma(\frac{5}{4})}{\Gamma(\frac{7}{4})}\right) =-0.06319... 
\eeqa
where $A$ is Glaisher's constant $A=1.28242...$. In the FB theory, the branch-point twist field has vacuum expectation value
\beq
	{}_n\bra\vac|{\cal T}|\vac\ket_n = 
    (2 \pi)^{\frc{n-1}2}
    \lt(\frac{\Gamma(n)}{n^n}\rt)^{\frac{1}{2}} 
    \lt(\frc m2\rt)^{2\Delta_n}
	\exp\int_0^\infty \frc{dt}t e^{-t}
	\lt(\frc{n\coth\frc t2 - \coth \frc t{2n}}{2(1-e^{-t})} - 2\Delta_n\rt)
\eeq
where in $\Delta_n$ the central charge is $c=1$, and this gives
\beq 
	\frc{\mathcal{E}_{\rm shift}^{\rm FB}
    +\mathcal{E}_{\rm sat}^{\rm FB}}2 = \frc14 \log\frc 4\pi
    = 0.06039...
\eeq 
In both cases the expectation values may been obtained by employing the angular quantization approach developed in \cite{Luky1,Luky2}. The free Boson computation will be presented in detail elsewhere \cite{pre}.

Concerning the other terms in (\ref{sca1}) the integral correction in the third line is in fact zero for the FB theory as all form factors vanish in the limit $n\rightarrow \frac{1}{2}$. On the other hand, in the FM theory with mass $m$, this correction is exactly given by
\beq
	\frc1{(2\pi)^2} \int_{-\infty}^\infty d\theta\,\text{sech}\frac{\theta}{2}\text{sech}\frac{3\theta}{2}\,
	K_0(2mr\cosh(\theta/2)).
\eeq
We thus have overall
\beqa
\lefteqn{\mathcal{E}^{\perp,\,\rm FM} = -\frc 18 \log(m\varep)
-0.06319+\log(C_1)} && \n &&-\frac{2}{3\sqrt{3}\pi} K_0(\sqrt{3} m r)+\frc1{(2\pi)^2} \int_{-\infty}^\infty d\theta\,\text{sech}\frac{\theta}{2}\text{sech}\frac{3\theta}{2}\,
	K_0(2mr\cosh(\theta/2))+\cdots
\eeqa
and 
\beq 
\mathcal{E}^{\perp,\,\rm FB}=-\frc 14 \log(m\varep)+0.06039{+\log(C_1)}-\frac{2}{3\sqrt{3}\pi} K_0(\sqrt{3} m r)+ \cdots
\eeq 

\noindent {\bf Acknowledgment:} The authors would like to thank the Galileo Galilei Institute for Theoretical Physics, (Arcetri, Florence) for their hospitality and financial support during the scientific program on ``Statistical Mechanics, Integrability and Combinatorics" held in May--June (2015).  OBF's work was supported by the ``Fonds de Recherche du Qu\'ebec, Nature et Technologie" (FRQNT).

\end{document}